\documentclass[epj]{svjour}
\usepackage{latexsym}
\usepackage{graphics}
\usepackage{epsfig}
\def\bea{\begin{eqnarray}}
\def\eea{\end{eqnarray}}
\def\beq{\begin{equation}}
\def\eeq{\end{equation}}
\def\nn{\nonumber}
\newcommand{\Lc}{{\Lambda}_c}
\newcommand{\Sc}{{\Sigma}_c}
\newcommand{\La}{{\Lambda}}
\newcommand{\Si}{{\Sigma}}
\newcommand{\ccc}{{\,\cdot\cdot\cdot\,}}
\newcommand{\Yc}{{Y}_c}
\begin{document}
\title{Scattering of charmed baryons on nucleons}
\author{J. Haidenbauer\inst{1}\thanks{j.haidenbauer@fz-juelich.de}
\and G.~Krein\inst{2}\thanks{gkrein@ift.unesp.br}
} 
\institute{Institute for Advanced Simulation, Institut f\"ur Kernphysik, 
and J\"ulich Center for Hadron Physics,
Forschungszentrum J\"ulich, D-52425 J\"ulich, Germany  \and 
Instituto de F\'{\i}sica Te\'orica, Universidade Estadual 
Paulista \\
Rua Dr. Bento Teobaldo Ferraz, 271 - Bloco II, 01140-070, 
S\~ao Paulo, SP, Brazil}
\date{Received: date / Revised version: date}
%
\abstract{
Chiral effective field theory is utilized for extrapolating results on the 
$\Lambda_c N$ interaction, obtained in lattice QCD at unphysical (large) quark 
masses, to the physical point. 
The pion-mass dependence of the components that constitute the $\Lambda_c N$ potential 
up to next-to-leading order (pion-exchange diagrams and four-baryon contact terms)
is fixed by information from lattice QCD simulations. No recourse to SU(3) or SU(4) 
flavor symmetry is made. 
It is found that the results of the HAL QCD Collaboration for quark masses
corresponding to $m_\pi = 410$--$570$ MeV imply a moderately attractive $\Lambda_c N$ 
interaction at $m_\pi = 138$ MeV with scattering lengths of $a\approx -1$ fm for the
$^1\!S_0$ as well as the $^3\!S_1$ partial waves.  
For such an interaction the existence of a charmed counterpart of the hypertriton 
seems unlikely but four- and/or five-baryons systems with a $\Lambda_c$ baryon could be 
indeed bound.
} 
\maketitle
\section{Introduction}
\label{sec:intro}

The dynamics of hadrons with different flavor degrees of freedom provides different windows for 
our understanding of the underlying theory of strong interaction, quantum chromodynamics (QCD). 
With regard to hadrons with charm, so far spectroscopy has been the most visible 
and definitely by far the most interesting branch of research. Indeed, the 
large number of structures observed in experiments at energies above the open charm 
production threshold provides a challenge for our standard (but obviously naive) picture 
that mesons are composed out of quark-antiquark pairs and baryons out
of three quarks. See Refs.~\cite{Guo:2017,Lebed:2016hpi,Chen:2016,Esposito:2016} 
for recent overviews and discussions of these structures, commonly
referred to as X, Y and Z states. 

Some proposals for experiments at sites such as J-PARC \cite{JPARC} 
and FAIR \cite{CBM,Wiedner:2011} aim at exploring also other aspects of charm physics. 
Specifically, so-called charm factories would allow 
the targeted production of charmed hadrons like the $D$-meson or the $\Lc$ and 
$\Sc$ hyperons and a study of their interaction with ordinary hadrons. 
The expectation of possible experiments in the not so far future has triggered a variety 
of theoretical investigations. In particular, it has led to a renewed interest in the 
interaction of the $\Lc$ with nucleons and with nuclei over the last few years  
\cite{Liu:2011xc,Huang:2013,Gal:2014,Garcilazo:2015qha,Maeda:2015hxa,Shyam:2016uxa,Ohtani:2017wdc}.
Those studies join the ranks of a long history of speculations about bound nuclear systems 
involving the $\Lc$, the lightest charmed 
baryon~\cite{Dover:1977jw,Bando:1981ti,Bando:1983yt,Gibson:1983zw,Bunyatov:1992in,Tsushima:2002ua,Tsushima:2003dd,Kopeliovich:2007kd} {\textemdash} see also the recent reviews~\cite{Hosaka:2016ypm,Krein:2017usp}. 
Indeed, in most of the investigations so far, the $\Yc N$ interaction 
($\Yc = \Lambda_c, \Sigma_c$), derived within the meson-exchange 
framework~\cite{Liu:2011xc,Dover:1977jw,Bando:1981ti,Bando:1983yt,Gibson:1983zw} or in 
the constituent quark model~\cite{Maeda:2015hxa,Froemel:2004ea}, turns out to be 
strongly attractive.  
 
Interestingly, quite the opposite picture emerged from recent (2+1)-flavor lattice QCD (LQCD) 
simulations by the HAL QCD Collaboration~\cite{Miyamoto:2016,Miyamoto:2016A,Miyamoto:2017}. 
Pertinent calculations, performed for unphysical quark masses corresponding to pion masses\footnote{The 
Gell-Mann-Oakes-Renner relation states that the squared pion mass is proportional to the average 
light quark mass. Therefore, the notions ``quark mass'' and ``pion mass'' are used synonymously.}
of $m_\pi = 410 - 700$ MeV, 
suggest that the $\Lc N$ and $\Sc N$ interactions could be much less attractive
than predicted by the phenomenological potentials mentioned above.
While initial preliminary studies \cite{Miyamoto:2016,Miyamoto:2016A} indicated
an extremely weak $\Lc N$ interaction, the recently published final LQCD results
\cite{Miyamoto:2017} rectify that conjecture and imply a somewhat stronger, 
though still only moderately attractive $\Lc N$ interaction\footnote{We note that the 
method employed in these works and also other methods are
presently under discussion in the LQCD 
community~\cite{Aoki:2017,Yamazaki:2017,Davoudi:2017,Aoki:2017A}.}. 

In the present work we provide predictions for the $\Lc N$ interaction at the physical point 
based on the LQCD simulations by the HAL QCD Collaboration \cite{Miyamoto:2017}. 
The extrapolation of the LQCD results, available for $m_\pi = 410 - 700$ MeV, 
to $m_\pi = 138$ MeV is performed within the framework of chiral effective field 
theory (EFT)~\cite{Epelbaum:2005,Epelbaum:2008ga,Machleidt:2011zz}. 
Thereby we follow a strategy that has been already employed by us in the past in the 
analysis of other LQCD results on baryon-baryon interactions in the strangeness sector, 
notably the $\Lambda\Lambda$~\cite{Haidenbauer:2011a}, $\Xi\Xi$ \cite{Haidenbauer:2011b}, 
and $\Omega\Omega$ \cite{Haidenbauer:2017} systems: 
At first chiral EFT is utilized to establish a $\Lc N$ potential for baryon and meson 
masses that correspond to those in the lattice simulation. In particular, open parameters 
are determined by a fit to pertinent LQCD results (phase shifts, scattering lengths). 
Then the potential is extrapolated to the physical point. Thereby the pion-mass 
dependence of the ingredients (pseudoscalar-meson exchange, four-baryon 
contact terms) is taken into account explicitly, and in line with chiral EFT. 

The paper is structured as follows: In Sect.~2 we provide an outline of the
employed formalism. 
Results for $\Lc N$ phase shifts (for the $^1\!S_0$ and $^3\!S_1$ partial waves)
are reported in Sect.~3, for pion masses corresponding to those in the
LQCD simulation and for the physical value, $m_\pi = 138$ MeV. 
The corresponding scattering lengths are evaluated too and turn out to be in
the order of $a\approx -1$~fm. Finally, consequences of our results for the 
existence of bound $\Lc$ hypernuclei are discussed. 
The paper closes with a short summary. 

\section{Formalism}
\label{sec:formalism}

The $Y_c N$ interaction is constructed by using chiral EFT as guideline.
Thereby we follow closely our application of this scheme to the $\Lambda N$ and
$\Sigma N$ systems in Refs.~\cite{Polinder,Haidenbauer:2013} where corresponding 
potentials have been obtained up to NLO in the Weinberg power counting  
\cite{Epelbaum:2008ga,Machleidt:2011zz}.
In this framework the potential is given in terms of pion exchanges 
and a series of contact interactions with an increasing number of derivatives. 
The latter represent the short-range part of the baryon-baryon force and are
parameterized by low-energy constants (LECs), that need to be fixed in a fit 
to data.
Both classes of contributions depend on the quark mass (or, equivalently, the pion mass).
At LO the only quark-mass dependence of the $\Yc N$ potential is through the 
pion mass that appears in the propagator of the pion-exchange potential, cf. below. 
However, at NLO the contact terms as well as the pion coupling constants depend on the 
pion mass. For details, we refer to
Refs.~\cite{Beane:2002vs,Beane03,Epe02,Epe02a,Baru:2015,Baru:2016},
where the quark mass dependence of the $NN$ interaction has been investigated;
see also Ref.~\cite{Petschauer:2013}. 

Let us start by introducing the contact interaction that we employ. For the partial waves 
considered in the present study ($^1\!S_0$, $^3\!S_1$-$^3\!D_1$) it is given by 
\begin{eqnarray}
V(^1\!S_0) &=& {\tilde{C}}_{^1\!S_0} + \tilde{D}_{^1\!S_0} m^2_\pi \nn \\
&& + \, ({C}_{^1\!S_0} 
+ {D}_{^1\!S_0} m^2_\pi)\, ({p}^2+{p}'^2) \ , \nonumber \\
V(^3\!S_1) &=& {\tilde{C}}_{^3\!S_1} + \tilde{D}_{^3\!S_1} m^2_\pi \nn \\
&& + \, ({C}_{^3\!S_1} \,
+ {D}_{^3\!S_1} m^2_\pi)\, ({p}^2+{p}'^2) \ , \nonumber \\
V(^3\!D_1 -\, ^3\!S_1) &=& {C}_{\varepsilon_1}\, {p'}^2 \ , \nonumber \\
V(^3\!S_1 -\, ^3\!D_1) &=& {C}_{\varepsilon_1}\, {p}^2 \ , 
\label{LEC}
\end{eqnarray}
with $p = |{\bf p}\,|$ and ${p}' = |{\bf p}\,'|$  being the initial and final 
center-of-mass (c.m.) momenta in the $\Lc N$ or $\Sc N$ systems. 
The quantities ${\tilde{C}}_{i}$, ${\tilde{D}}_{i}$, ${C}_{i}$, ${D}_{i}$ are the 
aforementioned LECs that need to be fixed by a fit to lattice data (phase shifts).
The ansatz Eq.~(\ref{LEC}) is motivated by the corresponding expression in the standard Weinberg
counting up to NLO \cite{Epe02,Petschauer:2013} but differs from it by the terms proportional to 
$m^2_\pi\, ({p}^2+{p}'^2)$ which are nominally of higher order. Nevertheless, we include these 
terms because they allow us to obtain an optimal description of the LQCD results at $m_\pi=410$ MeV 
as well as at $570$ MeV and, thereby, enable us a better constrained extrapolation to lower pion 
masses. Contrary to our study of the $YN$ interaction in Refs.~\cite{Polinder,Haidenbauer:2013},
here we do not impose SU(3) (or SU(4)) flavor symmetry. 
In any case, given that there is no explicit information on the $\Sc N$ channel from LQCD in 
Ref.~\cite{Miyamoto:2017} and the $\Lc N$ interaction is determined as an effective 
single-channel potential, we treat the $\Lc N$ interaction likewise as an effective 
single-channel problem. 

The contribution of pion exchange to the $\Yc N$ potential is given by 
\begin{equation}
V^{OPE}_{{B N\to B' N}} =-f_{{B B'\pi}}f_{{NN\pi}}
\frac{\left({\bf \sigma}_1\cdot {\bf q} \right)
\left({\bf \sigma}_2\cdot {\bf q} \right)}{ {\bf q}^{\,2}+m_{\pi}^2} 
\mathcal{I}_{{B N\to B' N}}\ ,
\label{OPE}
\end{equation}
where $B,B'$ stand for $\Lc$ and/or $\Sc$, 
${\bf q}$ is the transferred momentum, ${\bf q} = {\bf p'} - {\bf p}$,
and $\cal{I}$ is a pertinent isospin factor \cite{Polinder}.  
As already mentioned, we do not assume the validity of SU(4) flavor symmetry in the 
present study. This concerns also the coupling constants $f_{{B B'\pi}}$. 
The $\Lc\Sc\pi$ coupling constant can be determined from the experimentally 
known $\Sc \to \Lc\pi$ decay rate, see Refs.~\cite{Albertus:2005,Can:2016}. 
With regard to the $\Sc\Sc\pi$ coupling constant we resort to LQCD results~\cite{Alexandrou:2016}.
Besides their value at the physical point, we need also 
the $m_\pi$ dependence of the $\Lc\Sc\pi$ and $\Sc\Sc\pi$ 
coupling constants as well as the one for the $NN\pi$ vertex,
\begin{equation}
f_{BB'\pi} (m^2_\pi) = \frac{g_A^{BB'}(m^2_\pi)}{2\,F_\pi (m^2_\pi) } \ .
\label{fpi}
\end{equation}
LQCD results for the $m^2_\pi$ dependence of the pion decay constant $F_\pi$ are
readily available in the literature, e.g. in Ref.~\cite{Durr:2013}. From that 
reference, one deduces the values 
$F_\pi \approx 112$ MeV at $m_\pi = 410$ MeV,
$F_\pi \approx 129$ MeV at $m_\pi = 570$ MeV, and
$F_\pi \approx 141$ MeV at $m_\pi = 700$ MeV. To obtain the 
latter value, a linear $m^2_\pi$ dependence of $F_\pi$ has been assumed, 
as suggested by {Fig.~5} in Ref.~\cite{Durr:2013}.
LQCD results for the $m^2_\pi$ dependence of the axial-vector strengths $g^{BB}_A$ 
can be found in Ref.~\cite{Alexandrou:2016} though only up to 
$m_\pi = 500$ MeV. For $g^{\Sc\Sc}_A$ a rather moderate increase with $m_\pi$ is 
suggested by LQCD~\cite{Alexandrou:2016}. $g^{NN}_A$ is found to be practically 
independent of $m_\pi$, though, unfortunately, the lattice results do not match 
well with the known value at the physical point. There is no information on
the variation of the $\Lc\Sc\pi$ vertex with $m_\pi$. Because of these reasons 
we neglect the dependence of the $g_A$'s on $m_\pi$ in our calculation and  
assume that $g_{A}^{BB'}\equiv g_{A}^{BB'}(m_\pi=138~{\rm MeV})$.
Specifically, we use $g_A^{NN}=1.27$~\cite{PDG},
$g_A^{\Sc\Sc}=0.71$~\cite{Alexandrou:2016} 
and $g_A^{\Lc\Sc}=0.74$~\cite{Albertus:2005,Can:2016}. 
Note that the by far strongest dependence of $f_{BB'\pi}$ on $m^2_\pi$ 
comes via $F_\pi (m^2_\pi)$ \cite{Durr:2013} and this circumstance is 
adequately taken into account in our calculation.

Since $f_{\Lc\Lc\pi} \equiv 0$ under the assumption that isospin is
conserved, there is no one-pion exchange contribution to
the $\Lc N \to \Lc N$ potential. However, we include the coupling
of $\Lc N$ to $\Sc N$ via pion exchange, which is known to play an important
role in case of the $\Lambda N$ and $\Sigma N$ systems \cite{Haidenbauer:2013}.
The resulting effective two-pion exchange contribution to the
$\Lc N$ potential is generated by solving a coupled-channel 
Lippmann-Schwinger (LS) equation, see below. In this context let us note that the pertinent 
contribution arises anyway at NLO, even in a single-channel treatment \cite{Beane:2003}. 
In principle, at NLO there are further contributions from two-pion exchange
\cite{Haidenbauer:2013}. However, in the present study we omit those for simplicity 
reasons and assume that they are effectively absorbed into 
the contact terms. Furthermore, contributions from $\eta$ and $D-$meson 
exchanges that would arise under the assumption of SU(3) (or SU(4)) symmetry,
are likewise delegated to the contact interactions. 

As already mentioned above, in line with the analysis of the LQCD data in
Ref.~\cite{Miyamoto:2017}, we treat the $\Lc N$ interaction as an effective
single-channel problem. Possible effects from the interaction in the
$\Sc N \to \Sc N$ channel are thereby absorbed into the LECs of the 
$\Lc N$ potential.
A further issue that we have ignored is the role of heavy quark spin 
symmetry~\cite{Liu:2011xc,Lu:2017}. Indeed the $\Sc N$ and ${\Sc^*} N$ thresholds 
are just about $65$ MeV apart \cite{PDG} so that the coupling between those systems 
could be important \cite{Liu:2011xc}. However, since already the ${\Sc} N$ channel is 
not explicitly considered, there is no point in including ${\Sc^*} N$ in our analysis.
If there are any effects of it, these are likewise absorbed into the $\Lc N$ LECs. 
Anyway, since $M_{\Sc^*}-M_{\Lc} = {234}$ MeV, 
there should be less influence on the $\Lc N$ amplitude. Indeed, because of the larger 
mass difference $M_{\Sc}-M_{\Lc} \approx 167$~MeV, as compared to $M_{\Sigma}-M_{\Lambda} 
\approx 78$~MeV, one would expect that even the channel coupling to ${\Sc} N$ plays a 
less important role for the charm sector than in the strangeness sector. 

The reaction amplitudes are obtained from the solution of a coupled-channel 
LS equation for the interaction potentials, which is
given in partial-wave projected form by 
\begin{eqnarray}
T^{\nu'\nu,J}_{\rho'\rho}(p',p;\sqrt{s})&=& V^{\nu'\nu,J}_{\rho'\rho}(p',p) \nonumber\\
&& + \sum_{{\rho''},\nu''}\int_0^\infty \frac{dp''p''^2}{(2\pi)^3} \, 
V^{\nu'\nu'',J}_{\rho'\rho''}(p',p'') \nn \\
&&\times\, \frac{2\mu_{\rho''}}{p^2_{\rho}-p''^2+i\eta}
T^{\nu''\nu,J}_{\rho''\rho}(p'',p;\sqrt{s}).
\label{LS}
\end{eqnarray}
Here, the label $\rho$ indicates the particle channels and the label $\nu$ the partial wave.
$\mu_\rho$ is the pertinent reduced mass. The on-shell momentum in the intermediate state,
$p_{\rho}$, is defined by 
$\sqrt{s}=\sqrt{m^2_{B_{1,\rho}}+p_{\rho}^2}+\sqrt{m^2_{B_{2,\rho}}+p_{\rho}^2}$.
Following the practice of the HAL QCD Collaboration~\cite{Miyamoto:2017},
phase shifts will be given as functions of the kinetic energy in the $\Lc N$ 
c.m. frame, $E = \sqrt{s} - M_{\Lc}- M_N$. 

For baryon-baryon potentials constructed along the lines of chiral EFT
(cf. Eqs. (\ref{LEC}) and (\ref{OPE})) a regularization is required when solving the
LS equation (\ref{LS}) \cite{Epelbaum:2005,Machleidt:2011zz}.
Ideally, the resulting reaction amplitude $T$ should be completely independent of
the employed regularization scheme. In practise, there is still a dispute about how
regularization should be done in the application of chiral EFT to $NN$ scattering
(and accordingly in the $YN$ or $Y_cN$ case) to satisfy all formal aspects from a
field theory point of view and there is no generally accepted scheme,
cf. Refs.~\cite{Epelbaum:2005,Machleidt:2011zz} and references therein.
A commonly accepted procedure is the introduction of a cutoff into the
Lippmann-Schwinger equation or (equivalently) to the potential.
The controversial issue is, however, how one should then proceed in
detail in order to achieve the desired cutoff independence of the results
see, e.g.\ \cite{Lepage:1997,Epe05,Nogga:2005,Mach:2013,Phillips:2013,Epe15,Epelbaum:2017}.

In the present work we refrain from elaborating on this still open question.
Rather we follow strictly the procedure used in the application of chiral EFT to $NN$ scattering 
by Epelbaum and collaborators \cite{Epe05} and in our $YN$ studies \cite{Haidenbauer:2013}, 
where the potentials in the LS equation are cut off in momentum space by multiplication 
with a regulator function, $f(p',p) = \exp\left[-\left(p'^4+p^4\right)/\Lambda^4\right]$,
so that the high-momentum components of the baryon and pseudoscalar meson fields are
removed. The cutoff parameter $\Lambda$ in the regulator is typically in the order of
$\Lambda \approx 500$ MeV \cite{Epelbaum:2008ga,Machleidt:2011zz}. It is kept finite in the
calculation. Approximate cutoff independence is achieved by going to higher orders
in the perturbative expansion of the potential where the sucessively
arising contact terms allow one to absorb/remove the cutoff dependence
more and more efficiently \cite{Epe15}.
The actual values we employ for the cut-off are $\Lambda =500$-$600$ MeV,
in line with the range that yielded optimal and stable results in our NLO study
of the $\La N$ and $\Si N$ interactions \cite{Haidenbauer:2013}.
The variations of the results with the cutoff, i.e. the residual cutoff dependence,
reflect uncertainties that will be indicated by bands in the plots we show in the
next section.

In the analysis of the LQCD simulations we follow closely the strategy
of our previous works in Refs.~\cite{Haidenbauer:2011a,Haidenbauer:2011b,Haidenbauer:2017}:
(i) The LECs, i.e. the only free parameters in the potential, are determined by
a fit to LQCD results (phase shifts) employing the inherent baryon and meson masses 
of the lattice simulation;
(ii) Results at the physical point are obtained via a calculation in which
the pertinent physical masses of the mesons are substituted in the evaluation of
the potential and those of the baryons in the baryon-baryon propagators appearing
in the LS equation.
The baryon masses corresponding to the LQCD simulations at $m_\pi= 410$ and 
$570$~MeV are taken from Ref.~\cite{Miyamoto:2017}.
For the calculation at the physical point we use the masses from the 
PDG \cite{PDG}, i.e. $M_{\Lc} = 2286.5$ MeV, $M_{\Sc} = 2455$ MeV. 

Alternative ways to implement the quark mass dependence of the nuclear forces can
be found in Refs.~\cite{Baru:2015,Berengut:2013}.

\begin{table*}
\renewcommand{\arraystretch}{1.4}
\centering
\caption{Low-energy constants employed in the present work. 
$\tilde C_{^1\!S_0} \equiv {\tilde{C}}_{^1\!S_0} + \tilde{D}_{^1\!S_0} m^2_\pi$ 
at the specified pion mass, etc., see Eq.~(\ref{LEC}). The values
for $\tilde C$ are in 10$^4$~GeV$^{-2}$, those for $C$ in 10$^4$~GeV$^{-4}$.
}
\label{tab:lec} 
\vskip 0.2cm
\begin{center}
\begin{tabular}{||l||c|c||c|c||c|c||}
\hline
\hline
& \multicolumn{2}{c||}{$570$ MeV} & \multicolumn{2}{c||}{$410$ MeV} & \multicolumn{2}{c||}{$138$ MeV} \\
\hline
\hline
{$\Lambda = 500$ MeV} 
& $\tilde C$ & $C$ & $\tilde C$ & $C$ & $\tilde C$ & $C$ \\
\hline
$^1S_0$ &  -0.008576 & -0.006105  & -0.01758 & -0.01362  & -0.02615 & -0.02077 \\
$^3S_1$ &  0.03917   & 0.04569    & 0.1409   & 0.1270    & 0.2377   & 0.2043  \\
$^3D-{}^3S_1$ &      &-0.1136     &          & -0.1136   &          & -0.1136 \\
\hline
\hline
{$\Lambda = 600$ MeV} 
& $\tilde C$ & $C$ & $\tilde C$ & $C$ & $\tilde C$ & $C$ \\
\hline
$^1S_0$ &   -0.008026 &  0.001564  &  -0.01393 & -0.002252  &  -0.01954 & -0.005879 \\
$^3S_1$ &    0.04304  &0.07553  & 0.1619  &0.1688   & 0.2749  & 0.2574 \\
$^3D-{}^3S_1$ &      &-0.1344     &          & -0.1344   &          & -0.1344 \\
\hline
\hline
\end{tabular}
\end{center}
\end{table*}
\renewcommand{\arraystretch}{1.0}
 
\section{Results}
\label{sec:results}

LQCD results for phase shifts are available for the $\Lc N$ $^1\!S_0$ and $^3\!S_1$ 
partial waves for $m_\pi = 410,\, 570,\, 700$ MeV \cite{Miyamoto:2017}. 
We determine the LECs of the contact interaction, cf. Eq.~(\ref{LEC}),
by a fit to the lattice data at the two lower pion masses. Specifically, 
the pion-mass dependence exhibited by the LQCD simulation is exploited to determine 
the LECs $\tilde D_i$ and $D_i$ that encode the pion-mass dependence of the contact 
interaction. 
The fits are done to the phase shifts, generated from the parameterized version
of the $\Lc N$ potentials provided in Ref.~\cite{Miyamoto:2017}, for energies 
up to $30$~MeV. 
Alternative fits taking into account the HAL QCD results up
to $50$~MeV were performed too. In this context it should be noted that applications 
of chiral EFT to $NN$ scattering and specifically to the $^1\!S_0$ $np$ state reveal 
that NLO interactions are expected to provide quantitative results up to roughly 
$50$ MeV \cite{Epelbaum:2008ga,Machleidt:2011zz}. 
The actual values of the LECs at the fitted pion masses and
at the physical value are summarized in Table~\ref{tab:lec}. 
 
\begin{figure*}[t]
\begin{center}
\includegraphics[height=80mm,angle=-90]{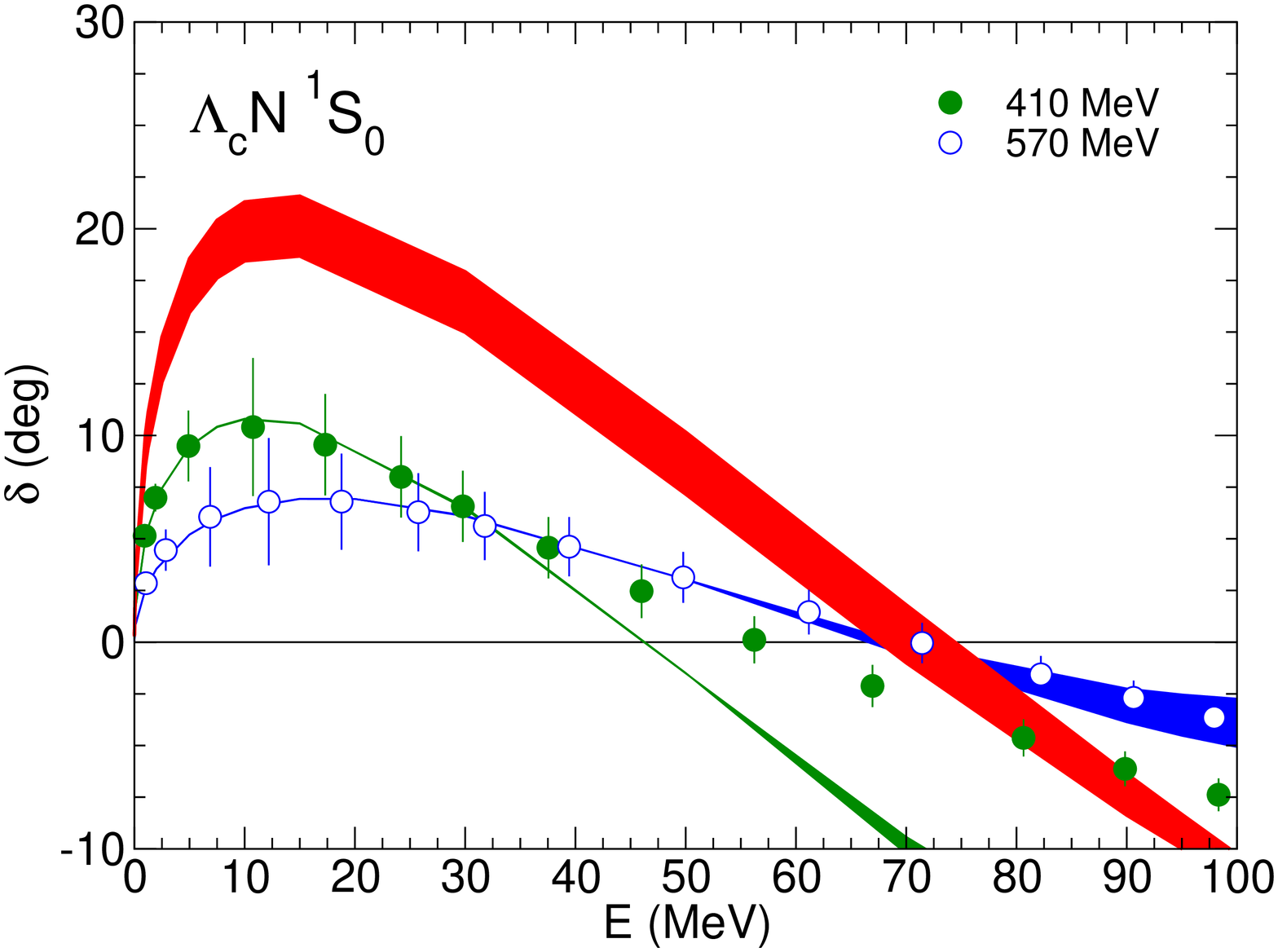} 
\includegraphics[height=80mm,angle=-90]{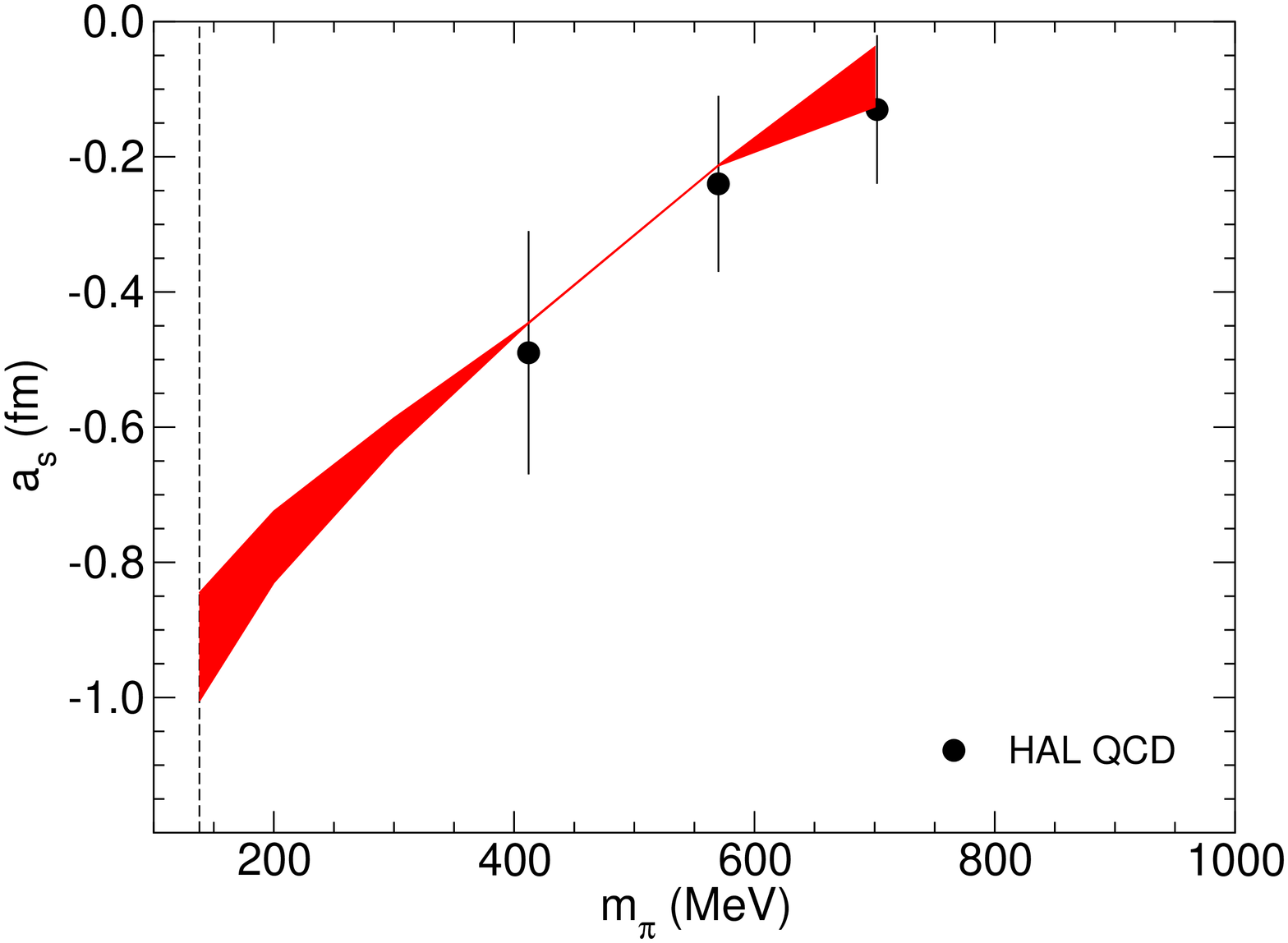}
\caption{Results for the $\Lc N$ $^1\!S_0$ partial wave. 
Left: Phase shifts of the HAL QCD Collaboration \cite{Miyamoto:2017} 
at $m_\pi = 570$ MeV (blue open circles) and $410$ MeV (green filled circles), 
as a function of the c.m. kinetic energy, together with our fits (lines/bands).  
The broader (red) band is the prediction for $m_\pi = 138$ MeV. 
Right: Dependence of the $^1\!S_0$ scattering length on the pion mass. 
The bands represent the cutoff variation $\Lambda = 500-600$ MeV, see text.
}
\label{ph1s0E}
\end{center}
\end{figure*}

Results for the $^1\!S_0$ partial wave are presented in Fig.~\ref{ph1s0E}.
The phase shifts for pion masses $570$, $410$ and $138$ MeV are shown
on the left side while the dependence of the scattering length on the
pion mass is depicted on the right side.
The bands represent the dependence of the results on variations of the 
cutoff $\Lambda$.
One can see that the lattice results at $m_\pi=410$~MeV are reproduced
quantitatively by our potential up to c.m. kinetic energies of around $40$~MeV, 
as expected for an NLO interaction, while those at $570$~MeV are remarkably 
well described over the whole energy range shown. 
In both cases the cutoff dependence is negligibly small at low energies so that
the bands are hardly visible.  
The phase shift obtained from our interaction when extrapolated to the physical 
point (red bands) do exhibit a noticeable but still rather moderate cutoff 
dependence. A maximum of around 20 degrees of the $^1\!S_0$ phase shift is 
predicted in this case.
The pion-mass dependence of the $^1\!S_0$ scattering length,
shown on the right-hand side of Fig.~\ref{ph1s0E}, is fairly smooth and almost
linear in $m_\pi$. Only close to the physical point a somewhat 
stronger $m_\pi$ dependence is visible. The value predicted at 
$138$ MeV is $a= -0.85 \ccc -1.0$~fm. 
 
\begin{figure*}[t]
\begin{center}
\includegraphics[height=80mm,angle=-90]{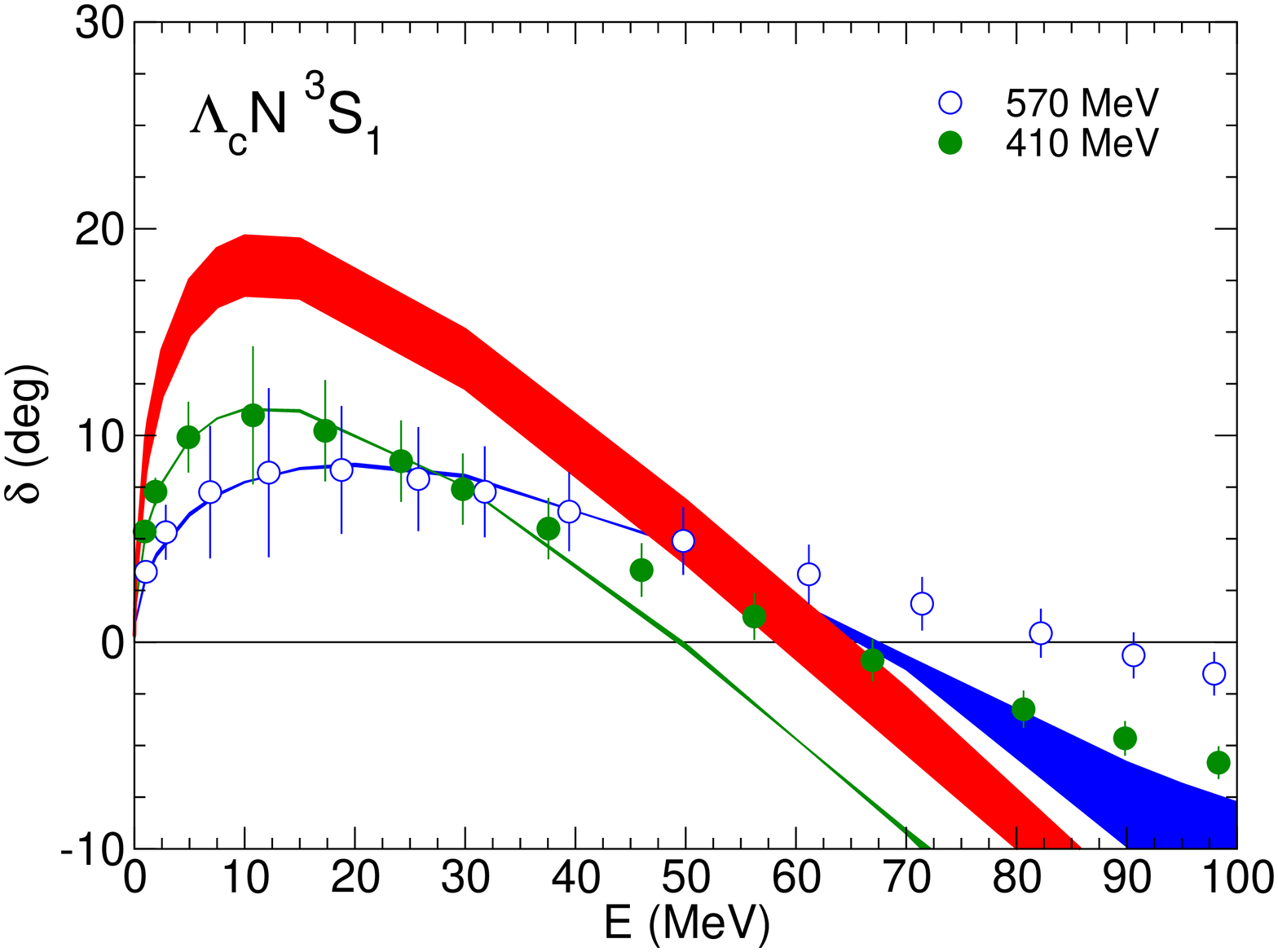}
\includegraphics[height=80mm,angle=-90]{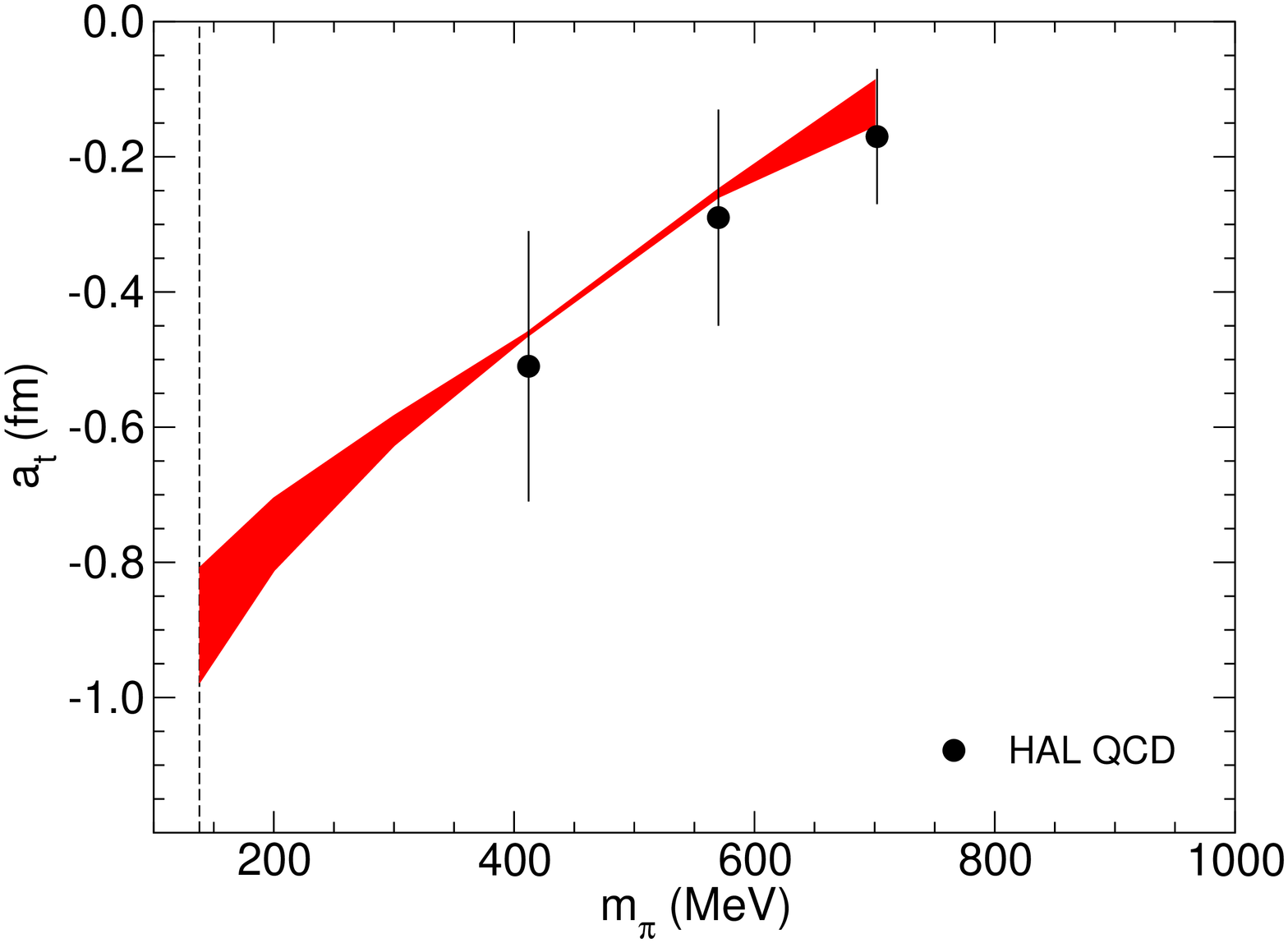}
\caption{Results for the $\Lc N$ $^3\!S_1$ partial wave.
Left: Phase shifts of the HAL QCD Collaboration \cite{Miyamoto:2017}    
at $m_\pi = 570$ MeV (blue open circles) and $410$ MeV (green filled circles),
as a function of the c.m. kinetic energy, together with our fits (lines/bands).
The broader (red) band is the prediction for $m_\pi = 138$ MeV.
Right: Dependence of the $^3\!S_1$ scattering length on the pion mass.
The bands represent the cutoff variation $\Lambda = 500-600$ MeV, see text.
}
\label{ph3s1E}
\end{center}
\end{figure*}

Results for the $^3\!S_1$ partial wave are presented in Fig.~\ref{ph3s1E}.
Again phase shifts as well as the pion-mass dependence of the scattering
length are shown. Those are very similar to the ones of the $^1\!S_0$ state,
even on a quantitative level. Indeed, it was already noted by the HAL
QCD Collaboration that the corresponding ($^1S_0$ and $^3S_1$) potentials 
they extracted are almost identical -- at 410 as well as at 570 MeV \cite{Miyamoto:2017}, 
and we see that this characteristic feature persists even in our extrapolation to the 
physical point. This is certainly a remarkable feature in view of the fact that in case 
of the $^3\!S_1$ partial wave there is a coupling to the $^3\!D_1$ induced by the
tensor force, and this coupling is taken into account in our analysis.
 
\begin{table*}
\renewcommand{\arraystretch}{1.2}
\centering
\caption{$\Lc N$ scattering lengths in the $^1\!S_0$ ($a_s$) and $^3\!S_1$ ($a_t$)
partial waves (in fm) for different pion masses. Predictions for the corresponding
effective range parameters $r$ (in fm) at the physical point are likewise given. 
}
\label{tab:a} 
\vskip 0.2cm
\begin{center}
\begin{tabular}{ l|l||c|c|c|c }
\hline
\hline
\multicolumn{2}{c||}{$m_\pi$} & $700$ MeV & $570$ MeV & $410$ MeV & $138$ MeV \\
\hline
\hline
$a_s$ & HAL QCD \cite{Miyamoto:2017} & $-0.13\pm 0.11$ & $-0.24\pm 0.13$ & $-0.49\pm 0.18$ & $-$ \\
        &our results    & $-0.04\ccc -0.13$ & $-0.21$ & $ -0.45 $ & $-0.85\ccc -1.00$ \\
$r_s$ &                              &                 & &        & $2.88\ccc 2.61$ \\
\hline
$a_t$ & HAL QCD \cite{Miyamoto:2017} & $\quad -0.17\pm 0.10\quad$ & $\quad -0.29\pm 0.16\quad$ 
& $\quad -0.51\pm 0.20\quad$ & $-$ \\
 & our results & $-0.09\ccc -0.15$ & $-0.25\ccc -0.26$ & $ -0.46\ccc -0.47 $ & $-0.81\ccc -0.98$ \\
$r_t$ &                              &                 & &                   & $3.50\ccc 3.15$ \\
\hline
\hline
\end{tabular}
\end{center}
\end{table*}
\renewcommand{\arraystretch}{1.0}
 
For completeness we have summarized the results for the $^1\!S_0$ and $^3\!S_1$
scattering lengths, $a_s$ and $a_t$, in Table~\ref{tab:a}.
From those numbers one can see that the variation in the extrapolated values due
to the employed regularization scheme is in the order of $0.2$ fm. 
Additional fits to the phase shifts where the considered energy range was extended 
up to $50$ MeV led to changes in the scattering lengths at the physical point of 
around $0.1$ fm, with a clear tendency to smaller values. 
Further exploratory fits carried out by us indicate that variations in the
scattering length of $\pm 0.05$~fm at $m_\pi = 410$~MeV amount to differences 
of about $\pm 0.10$~fm at $m_\pi = 138$ MeV. Combining these observations with
the uncertainty of $\pm 0.2$~fm given by the HAL QCD Collaboration for their 
result at $m_\pi = 410$ MeV suggests that the scattering lengths
at the physical point could be about $0.3$~fm larger, i.e. as large as $-1.3$~fm. 
Finally, for estimating possible effects from the $\Sc N$ channel we performed 
calculations with the pion-exchange contribution (\ref{OPE}) to the $\Sc N$ potential 
included. Adding its contribution within a coupled-channel framework (\ref{LS}) 
has a negligible influence on the results at $m_\pi = 410$ MeV and $570$ MeV, but 
leads to noticeable variations in the predictions for $m_\pi = 138$ MeV that
amount to roughly $0.3$~fm in the scattering lengths. 
Of course, a complete calculation has to include also a contact term in the $\Sc N$ 
channel, which could facilitate a reduction of the effect. 
Obviously, additional information on the $\Sc N$ phase shifts, as had been
provided in the preliminary study of the HAL QCD Collaboration \cite{Miyamoto:2016A},
and promised in Ref.~\cite{Miyamoto:2017} for the future,
would be helpful for reducing the uncertainty in the extrapolation. 
In any case, a more quantitative estimate of the overall uncertainty should
be attempted, once lattice data with better statistics are 
available and/or results for pion masses closer to the physical point. 
 
Clearly, and in line with the LQCD results for larger pion masses, we do not
get any $\Lc N$ bound states. 
However, what are the consequences of the results presented in the preceding
paragraphs for the possible existence of bound $\Lc$ nuclei? 
Let us look
at the strangeness sector and, specifically, at the lightest $\Lambda$ nucleus 
that is experimentally observed, namely the hypertriton $^3\!_\Lambda \rm H$.
The experimental value for the $^3\!_\Lambda \rm H$ binding energy is 
$(-2.354\pm 0.050)$ MeV,
which implies a separation energy for the $\Lambda$ of only $(0.13\pm 0.05)$ MeV.
Faddeev calculations of the coupled $\Lambda NN$--$\Sigma NN$ three-body systems
for realistic $YN$ potentials have been reported in Refs.~\cite{Nogga:2002,Haidenbauer:2007,Nogga:2013}.
Those suggest that, for $YN$ interactions which provide sufficient attraction 
so that the hypertriton is bound, the $\Lambda N$ scattering lengths are in the order of
$-2.9$ to $-2.6$ fm for the $^1\!S_0$ state and $-1.5$ to $-1.7$ fm for $^3\!S_1$.

The scattering lengths for the $\Lc N$ system at the physical point, deduced 
from the LQCD simulations of the HAL QCD Collaboration, are $-0.8$ to $-1.0$ fm 
for the $^1\!S_0$ as well as the $^3\!S_1$ states.  
Thus, in the case of the $^1\!S_0$ partial wave the value is considerably smaller
than its counterpart in the strangeness sector while for the $^3\!S_1$ channel the
difference is less dramatic. Since the average $\Lambda N$ (or $\Lc N$) potential 
that is relevant for the hypertriton is dominated by the spin-singlet channel, i.e. 
$V_{\Lambda N}\,(V_{\Lc N}) \propto \frac{3}{4}V_s+\frac{1}{4}V_t$, see Ref.~\cite{Gibson:1994} 
for details, it seems rather unlikely that an only moderately attractive singlet $\Lc N$ 
interaction can support the existence of a $^{\ \, 3}_{\Lc} \rm H$ bound state. 
Even effects due to a reduction of the kinetic energy associated with the $\Lc$ 
induced by the larger mass of the $\Lc$ as compared to the $\Lambda$, emphasized
in several works in the past \cite{Garcilazo:2015qha,Dover:1977jw,Gibson:1983zw},
cannot compensate for this large difference in the strength. 
Actually, in Ref.~\cite{Gibson:1983zw} an estimate for the binding energy of charmed
three- and four-body systems is provided based on an exact solution of 
corresponding scattering equations. Interestingly, Model~4 considered in that work
yields scattering lengths very similar to the ones of our analysis, namely
$a_s = -1.075$ fm and $a_t = -0.828$ fm (cf. Table I in that work). For this interaction 
no bound state is found for the $^{\ \, 3}_{\Lc} {\rm H}$ nucleus. However, the 4-body
systems $^{\ \, 4}_{\Lc} {\rm He}$ and $^{\ \, 4}_{\Lc} {\rm Li}$ could be already 
bound, though possibly only weakly, even when considering the additional
repulsive effect from the Coulomb force due to the charge of the 
$\Lc^+$ as discussed in Ref.~\cite{Gibson:1983zw}. 

With regard to recent few-body calculations, the $\Yc N$ interaction employed in 
Ref.~\cite{Maeda:2015hxa} is strongly attractive and leads already to $\Lc N$ 
bound states. The binding energies for $\Lc NN$
are then in the order of $20$ MeV. Those results do not allow us to draw any 
conclusions about what would happen for significantly less attractive $\Yc N$ 
interactions like the one we deduce from the analysis of the LQCD simulations 
by the HAL QCD Collaboration. 
In case of the three-body calculation presented in Ref.~\cite{Garcilazo:2015qha}, 
a charmed hypertriton with $J=3/2$ (and total isospin $I=0$) is predicted.
For that state the $\Lc N$ spin-triplet interaction is dominant~\cite{Miyagawa:1995}. 
However, since there is no information on the scattering length and effective
range parameters of the $\Lc N$ potential employed in
Ref.~\cite{Garcilazo:2015qha}, a proper assessment of that result is difficult.

\section{Summary}

In the present work, we have used the framework of chiral effective field theory 
to extrapolate lattice QCD results for the $\Lambda_c N$ interaction at $m_\pi=410 
- 570$~MeV by the HAL QCD Collaboration~\cite{Miyamoto:2017} to the physical point.
Thereby, we have followed a strategy employed previously in the 
strangeness sector~\cite{Haidenbauer:2011a,Haidenbauer:2011b,Haidenbauer:2017}.
However, contrary to the calculations performed in those works, no recourse to 
SU(3) or SU(4) flavor symmetry has been made in the present study.
Furthermore, in the present work the pion-mass dependence of all components that 
constitute the $\Lambda_c N$ potential up to next-to-leading order 
(pion-exchange diagrams and four-baryon contact terms)
are taken into account. Information from lattice QCD simulations is utilized to
implement these features.  
 
Our analysis of the HAL QCD results points to a moderately attractive $\Lambda_c N$ 
interaction at the physical point with scattering lengths of $a \approx -1$~fm for 
the $^1\!S_0$ as well as for the $^3\!S_1$ partial waves. 
Such an interaction leads to the possibility of bound four- and/or five-baryons 
systems with a $\Lambda_c$ baryon and presumably of heavier $\Lc$ hypernuclei. 
On the other hand, two-body ($\Lc N$) bound states as advocated in some recent
investigations based on phenomenological potentials~\cite{Liu:2011xc,Maeda:2015hxa} 
can be definitely excluded, even if one considers uncertainties in the extrapolation 
of the lattice results.
Also, the existence of a hypertriton-like $\Lambda_c NN$ three-body bound state
(with $J=1/2$) seems rather unlikely if one takes past investigations 
\cite{Gibson:1983zw} as benchmark. 

\vskip 0.2cm \noindent
{\bf Acknowledgments} \\
Work partially supported by Conselho Nacional de Desenvolvimento Cient\'{i}fico 
e Tecnol\'{o}gico (CNPq), Grant No. 305894/2009-9 (G.K.) and Grant No. 464898/2014-5 (G.K.) 
(INCT F\'{\i}sica  Nuclear e Applica\c{c}\~oes), Funda{\c c}\~{a}o de Amparo \`{a} Pesquisa 
do Estado de S\~{a}o Paulo (FAPESP), Grant No. 2013/01907-0 (G.K.)


\end{document}